\documentclass[superscriptaddress,secnumarabic,
amssymb,amsmath,nobibnotes,aps,prd,showkeys,showpacs,nofootinbib]{revtex4}%
\usepackage{graphicx}
\usepackage{epsf}
\usepackage{bm}
\usepackage{amsmath}
\usepackage{amsfonts}
\usepackage{amssymb}
\usepackage{epstopdf}
\usepackage{natbib}%
\providecommand{\U}[1]{\protect\rule{.1in}{.1in}}
\newcommand{\be}{\begin{equation}}
\newcommand{\ee}{\end{equation}}

\newcommand{\mincir}{\raise
-3.truept\hbox{\rlap{\hbox{$\sim$}}\raise4.truept\hbox{$<$}\ }}
\newcommand{\magcir}{\raise
-3.truept\hbox{\rlap{\hbox{$\sim$}}\raise4.truept\hbox{$>$}\ }}

\begin{document}
\title{Scalar field cosmology modified by the Generalized Uncertainty Principle}
\author{Andronikos Paliathanasis}
\email{paliathanasis@na.infn.it}
\affiliation{Dipartimento di Fisica, Universita' di Napoli Federico II, Complesso
Universitario di Monte S. Angelo, Via Cinthia, 9, I-80126 Naples, Italy}
\affiliation{Istituto Nazionale di Fisica Nucleare (INFN) Sez. di Napoli, Complesso
Universitario di Monte S. Angelo, Via Cinthia, 9, I-80126 Naples, Italy}
\author{Supriya Pan}
\email{span@research.jdvu.ac.in}
\affiliation{Department of Mathematics, Jadavpur University, Kolkata-700032, West Bengal, India}
\author{Souvik Pramanik}
\email{souvick.in@gmail.com}
\affiliation{Physics and Applied Mathematics Unit, Indian Statistical Institute, 203 B.T.
Road, Kolkata-700108, India}
\keywords{Cosmology; Dark energy; Scalar field; Generalized Uncertainty Principle}
\pacs{98.80.-k, 95.35.+d, 95.36.+x, 11.10.Ef}

\begin{abstract}
We consider quintessence scalar field cosmology in which the Lagrangian of the
scalar field is modified by the Generalized Uncertainty Principle. We show
that the perturbation terms which arise from the deformed algebra are
equivalent with the existence of a second scalar field, where the two fields
interact in the kinetic part. Moreover, we consider a spatially flat
Friedmann-Lema\^{\i}tre-Robertson-Walker spacetime (FLRW), and we derive the
gravitational field equations. We show that the modified equation of state
parameter $w_{GUP}$ can cross the phantom divide line; that is $w_{GUP}<-1$.
Furthermore, we derive the field equations in the dimensionless parameters,
the dynamical system which arises is a singular perturbation system in which
we study the existence of the fixed points in the slow manifold. Finally, we
perform numerical simulations for some well known models and we show that for
these models with the specific initial conditions, the parameter $w_{GUP}$
crosses the phantom barrier.

\end{abstract}
\maketitle

\section{Introduction}

\label{Introd}

Various approaches of quantum gravity, such as, String Theory (ST), Doubly
Special Relativity (DSR), Black Hole Physics (BHP) predict that there should
exist a minimum length scale of the order of the Planck length ($l_{PL}$), or,
in other words, the existence of a maximum energy scale in nature. This
prediction motivates to modify the Heisenberg's Uncertainty Principle to some
Generalized Uncertainty principle (GUP) in the context of the quantum theory
of gravity \cite{Maggiore}. The GUP model that has been proposed from String
Theory and Black Hole Physics contains the existence of minimum measurable
length, whereas DSR provides a GUP model which incorporates the existence of a
maximum observable momentum. This maximum momentum scale has been obtained in
the context of DSR theory for the consideration of the frame independent
invariant upper bound of energy scale into the theory \cite{Camelia}.
Interestingly, the above two scenarios, the minimum length scale and the
maximum momentum scale can be combined into a single deformed uncertainty
principle (DUP), which has the dual nature \cite{das}. However, the modified
Heisenberg's algebra that can be generated from this DUP relation is not
relativistically covariant. Hence, it is very difficult to apply the DUP in
any relativistic scenario.
So, here we mainly concentrate on the relativistically covariant modified
Heisenberg's algebra, which is constructed from the GUP model that
incorporates the minimum length scale only. Indeed, later on, we will see that
this modified algebra deforms the coordinate representation of the momentum
operator, whereas the position operator remains undeformed.

On the other hand, standard General Relativity (GR) is not enough to describe
the phenomena that recently appeared in astrophysics and cosmology, such as,
the late-time acceleration of the universe \cite{Riess1, Perlmutter1,
Spergel1, Komatsu2,planck,Percival1, Eisenstein1, Jain1}. An easy way to
explain the accelerated expansion of the universe is to consider an additional
fluid which has a negative equation of state parameter (dark energy); this new
fluid counteracts the gravitational force, and leads to the observed
accelerated expansion. The simplest dark energy candidate is the cosmological
constant $\Lambda$ leading to the $\Lambda$CDM cosmology \cite{Peebles1,
Padmanabhan1, Weinberg1}. However, this model has two serious problems: the
fine tuning problem \cite{Weinberg1, Carroll1}, and the coincidence problem
\cite{perivolaropoulos}. As a consequence, different dark energy models gave
rise into existence, and try to explain the recent accelerating phase of the
universe, such as, time-varying $\Lambda\left(  t\right)  $ cosmologies,
quintessence, k-essence, phantom and others (for instance see \cite{Lima92,
SolaBasil, Brookfield, Maleknejad, Starob, Faraoni, Ame10}).

Now, it is known from cosmic microwave background observations, that our
universe had gone through the inflationary phase at its early stage, and then
entered into the radiation, and matter dominated eras, in which the expansion
became decelerating. So, at the early stage of high energy regime, it was
accelerating, and, in this present low energy regime, it is again
accelerating. Quantum gravity effects played an important role at the early
universe, see for instance \cite{MB01,Kowa01,Kempf01,EGKS01,Amjad1,Amjad2}.
For instance, the mystery behind the origin of magnetic fields with $\mu$G
(microgauss) strength \cite{SFW86,AS87,K94} (having approximately the same
energy density as the cosmic microwave background radiation) observed in the
intergalactic scales, has some explanation in the context of GUP
\cite{Amjad04}. \ Furthermore, the existence of a minimum length, specifically
GUP, \textbf{ }prevent the complete evaporation of black holes \cite{Adler}.
In particular it has been shown that a black hole with temperature greater
than the ambient temperature should radiate photons, as well as, ordinary
particles until it reaches the Planck size, whereas in the Planck scale, it
stops radiating, and, its entropy reaches zero \cite{Adler} (for a review see
\cite{revBH}).

Thus, it will be very interesting to investigate whether the GUP corrected
modified theory can explain the present day accelerating phase, or, not.
Because, these theoretical approaches predict that the minimum length effects
are quite universal at any stage. Hence, they must have to describe the late
time accelerating phase in order to keep the consistency in the theory. Since,
it is known that the dark energy, is the proposed candidate to explain the
present day acceleration; so, we can study the consequences of the GUP
modification on the dark energy models.

We have already mentioned that the modified Heisenberg's algebra corresponding
to the GUP model that incorporates the existence of minimum length, modifies
the coordinate representation of momentum operator. As a consequence, the
dispersion relation of momentum become modified and finally the Klein-Gordon
equation deforms accordingly. From this modified Klein-Gordon equation one can
easily expect the deformation of a Lagrangian density in field theory. As, in
the context of dark energy we have a field theoretic model, namely, the
quintessence scalar field models \footnote{A quintessence scalar field has a
time varying equation of state (EoS) parameter $w_{\phi}\left(  z\right)  $
which is bounded as follows \ $\left\vert w_{\phi}\left(  z\right)
\right\vert \leq1$, and could describe the acceleration of the universe when
$w_{\phi}\left(  z\right)  <-\frac{1} {3}$.} \cite{Peebles1, Padmanabhan1,
Sahni, Lima2004, Basil2014}, so, we can modify it in light of GUP. Therefore,
in order to study the late-time acceleration of the universe, in this work, we
are interested on the scalar field models (quintessence) in which the
Lagrangian has been modified by the GUP. Note that, for this Lagrangian the
gravitation equations follows from the Einstein-Hilbert action with the usual
variational principle.

From the analysis of the recent data \cite{Spergel1, Komatsu2,planck} it
appears that the EoS parameter for the dark energy could have values
$w_{DE}\lesssim-1$. The limit $w_{DE}=-1$ is called phantom divide line or
phantom barrier. As, the usual quintessence scalar field model can not explain
this phase, so, our another motivation is to modify the scalar field model in
such a way that it can explain the recent phase. Indeed, here we will show the
modified model will drag a correction term into the effective equation of
state parameter $w_{GUP}\left(  z\right)  $, and it concludes that,
$w_{GUP}\left(  z\right)  $ can cross the above phantom barrier. The plan of
the paper is as follows.

In section \ref{GUPmd}, we give the basic definitions and properties of GUP
and we derive the modified Klein-Gordon equation. In section \ref{fequa}, we
consider a quintessence scalar field cosmology follows from GUP. We use
Lagrange multiplier and we write the gravitational action as that of two
scalar field cosmology with a mixed kinetic term. Furthermore, we consider a
spatially flat Friedmann-Lema\^{\i}tre-Robertson-Walker (FLRW) spacetime and
write the field equations and the modified EoS. In section \ref{cosmoevol}, we
write the field equations in the dimensionless variables, the dynamical system
which arises is a slow-fast dynamical system, where we study the fixed points
of the system in the slow manifold. Furthermore, we perform numerical
simulations for two well known scalar field potentials. Finally, in section
\ref{Discu}, we draw our conclusions.

\section{GUP-based modified model}

\label{GUPmd}

The GUP consistent with the existence of the minimum measurable length has the
structural form as follows
\begin{equation}
\Delta X_{i}\Delta P_{j}\geqslant\frac{\hbar}{2}[\delta_{ij}(1+\beta
P^{2})+2\beta P_{i}P_{j}]. \label{GUP}%
\end{equation}
The deformed Heisenberg algebra that can be found from the above GUP model
(\ref{GUP}) is given by \cite{Quesne2006,Vagenas,Kemph1,Kemph2},
\begin{equation}
\lbrack X_{i},P_{j}]=i\hbar\lbrack\delta_{ij}(1+\beta P^{2})+2\beta P_{i}%
P_{j}]. \label{xp-com}%
\end{equation}
Here $\beta$ is a parameter of deformation defined by $\beta={\beta_{0}%
}/{M_{Pl}^{2}c^{2}=\beta}_{0}\ell_{Pl}^{2}/2\hbar^{2}$, where $M_{Pl}$ is the
Planck mass, $\ell_{Pl}$ ($\approx10^{-35}~m)$ is the Planck length,
$M_{Pl}c^{2}$ ($\approx1.2~\times~10^{19}~GeV)$ is the Planck energy.

From the commutation relation (\ref{xp-com}), the coordinates representation
of the momentum operator can be modified to be $P_{i}=p_{i}(1+\beta p^{2}),$
by keeping $X_{i}=x_{i}$ undeformed. Here $(x,p)$ is the canonical
representation satisfying $[x_{i},p_{j}]=i\hbar\delta_{ij}$.

In the relativistic four vector form, the above commutation relation
(\ref{xp-com}) can be written as \cite{Quesne2006,Moayedi}
\begin{equation}
\lbrack X_{\mu},P_{\nu}]=-i\hbar\lbrack(1-\beta(\eta^{\alpha\gamma}P_{\alpha
}P_{\gamma}))\eta_{\mu\nu}-2\beta P_{\mu}P_{\nu}], \label{eqn3}%
\end{equation}
where the flat metric is given by $\eta_{\mu\nu}\equiv(1,-1,-1,-1)$. The
corresponding deformed operators in this case are \cite{Moayedi}%
\begin{equation}
P_{\mu}=p_{\mu}(1-\beta(\eta^{\alpha\gamma}p_{\alpha}p_{\gamma}))~,~~X_{\nu
}=x_{\nu}, \label{xp-form}%
\end{equation}
where$~p^{\mu}=i\hbar\frac{\partial}{\partial x_{\mu}},$ and $[x_{\mu},p_{\nu
}]=-i\hbar\eta_{\mu\nu}$.

For a spin-0 particle with rest mass $m$, the Klein-Gordon equation is
$\eta^{\mu\nu}P_{\mu}P_{\nu}\Psi-\left(  mc\right)  ^{2}\Psi=0$. By
substituting $P_{\mu}$ from (\ref{xp-form}), we have the modified Klein Gordon
equation has the following form%
\begin{equation}
\Delta\Psi+2\beta\hbar\Delta\left(  \Delta\Psi\right)  +\left(  \frac
{mc}{\hbar}\right)  ^{2}\Psi+O\left(  \beta^{2}\right)  =0, \label{kg.01}%
\end{equation}
where $\Delta=\frac{1}{\sqrt{-g}}\partial_{\mu}\left(  g^{\mu\nu}\sqrt
{-g}\partial_{\mu}\right)  $ is the Laplace operator, whereas for the metric
$g_{\mu\nu}\rightarrow\eta_{\mu\nu}$, $\Delta=\square=\partial_{\mu}%
\partial^{\mu}$. In contrary with the classical Klein-Gordon equation in
equation (\ref{kg.01}), there is the term $2\beta\hbar\Delta\left(  \Delta
\Psi\right)  $, which is the quantum correction term. Furthermore, we remark
that equation (\ref{kg.01}) is a fourth-order singular perturbed equation.

Equation (\ref{kg.01}) could arise from a variational principle, that is,
there exists a function $L\left(  \Psi,\Psi_{,\mu},\Psi_{,\mu\nu},...\right)
$, such that $\mathbf{E}\left(  L\right)  =0$, where $\mathbf{E~}$is the Euler
Lagrange operator. The Lagrangian of the fourth-order Klein-Gordon equation
has the following expression \cite{Moayedi},%
\begin{equation}
L\left(  \Psi,\mathcal{D}_{\sigma}\Psi\right)  =\frac{1}{2}g^{\mu\nu
}\mathcal{D}_{\mu}\Psi\mathcal{D}_{\nu}\Psi-\frac{1}{2}\left(  \frac{mc}%
{\hbar}\right)  ^{2}\Psi^{2}, \label{Lan2}%
\end{equation}
where the new operator $\mathcal{D}_{\mu}$ is $\mathcal{D}_{\mu}=\nabla_{\mu
}+\varepsilon\nabla_{\mu}\left(  \Delta\right)  $;$~\nabla_{\mu}$ is the
covariant derivative, i.e.$~\nabla_{\mu}\phi=\phi_{;\mu}$, and $\varepsilon
=\beta\hbar^{2}={\beta}_{0}\ell_{Pl}^{2}/2$.

\section{Gravitational action and field equations}

\label{fequa}

Let a four dimensional Riemannian manifold with metric $g_{\mu\nu}$ and
curvature $R$. We consider the following gravitational action%
\begin{equation}
S=S_{GR}+S_{GUP}+S_{m},
\end{equation}
where $S_{GR}=\int d^{4}x\sqrt{-g}R$ is the Einstein-Hilbert action,\ $S_{m}$
\ describes the matter components, and $S_{GUP}$ is the action of the scalar
field modified by the GUP, that is,
\begin{equation}
S_{GUP}=-\int\sqrt{-g}\left(  \frac{1}{2}g^{\mu\nu}\mathcal{D}_{\mu}%
\phi\mathcal{D}_{\nu}\phi-V(\phi)\right)  .
\end{equation}

Therefore, the gravitational action is\footnote{In contrary to the
Saez-Ballester scalar field model \cite{Saez}, in which the operator
$\mathcal{D}_{\mu}$, can be seen as $f\left(  \phi\right)  \nabla_{\mu}$,
where $\nabla_{\mu}$, is the covariant derivative, the new operator
$\mathcal{D}_{\mu}$ includes second-order derivatives. For the cosmological
viability of the Saez-Ballester model see \cite{fisika}.}:
\[
S=\int\sqrt{-g}~d^{4}x\left(  R-\frac{1}{2}g^{\mu\nu}\mathcal{D}_{\mu}%
\phi\mathcal{D}_{\nu}\phi+V(\phi)\right)  +S_{m}~,
\]
or, equivalently,
\begin{equation}
S=S_{GR}+S_{\phi}+S_{m}+\int\sqrt{-g}~d^{4}x\left(  -\varepsilon g^{\mu\nu
}\phi_{;(\mu}g^{ab}\phi_{;\left\vert ab\right\vert \nu)}\right)  ,
\label{ac.04}%
\end{equation}
where $S_{\phi}$ is the action of the \textquotedblleft
classical\textquotedblright\ scalar field \cite{Ame10}. Note that, the last
term of the above action (\ref{ac.04}) can be written equivalently as follows
\begin{equation}
S_{C}=\int\sqrt{-g}~d^{4}x\left(  \varepsilon g^{\mu\nu}\phi_{;\left(  \mu
\nu\right)  }g^{ab}\phi_{;ab}\right)  . \label{ac.05}%
\end{equation}

The action integral (\ref{ac.04}) provide us with a higher-order theory of
gravity in which the higher-order terms follow from the term $S_{C}$.

In order to simplify the calculations, we introduce a Lagrange
multiplier\footnote{Lagrange multipliers are useful to reduce to order of the
differential equations. However at the same time, the dimension of the
dependent variables is increased. For applications of the Lagrange multipliers
in high order theories of gravity, see for instance \cite{lan1,lan2,lan3}},
and, consider the new variable $\psi$ defined by $~\psi=g^{\mu\nu}\phi
_{;\mu\nu}$; therefore, incorporating the Lagrange multiplier $\lambda$ into
(\ref{ac.05}), and by solving the Lagrange's equation $\frac{\partial
S}{\partial\psi}=0$, we have $\lambda=-2\varepsilon\psi$. Hence, applying
integration by parts, equation (\ref{ac.05}) becomes
\begin{equation}
S_{C}=\int\sqrt{-g}~d^{4}x\left(  -2\varepsilon g^{ab}\phi_{,a}\psi
_{,b}-\varepsilon\psi^{2}\right)  . \label{ac.7b}%
\end{equation}

Therefore, the gravitational action (\ref{ac.04}) becomes
\begin{equation}
S=S_{GR}+S_{\phi}+S_{\psi}+S_{m}, \label{ac.08}%
\end{equation}
in which,~
\[
S_{\psi}=\int\sqrt{-g}~d^{4}x\left[  -2\varepsilon g^{ab}\phi_{,a}\psi
_{,b}-\varepsilon\psi^{2}\right]
\]
The last action describes a field which is coupled with the field $\phi$,
where the extra term follows from the perturbation effects of GUP.

Finally, from action (\ref{ac.08}), we have that the gravitation
equations$~$are
\begin{equation}
G_{\mu\nu}\equiv R_{\mu\nu}-\frac{1}{2}Rg_{\mu\nu}=8\pi GT_{\mu\nu}\left(
\phi,\psi\right)  +8\pi G\bar{T}_{\mu\nu}~, \label{einstein}%
\end{equation}
where $T_{\mu\nu}\left(  \phi,\psi\right)  $ is the energy-momentum tensor for
the two fields\footnote{For exact solutions of two scalar field models, see
\cite{Chervon1,TwoSF}, and for a quintom scenario with mixed kinetic term, see
\cite{Saridakis1}.} $\left\{  \phi,\psi\right\}  $, and, $\bar{T}_{\mu\nu}$ is
the energy- momentum of the matter component. We remark that the kinetic
metric which is defined by the two fields $\left\{  \phi,\psi\right\}  $ can
not be written in a diagonal form, since the \textquotedblleft
coordinate\textquotedblright\ transformation $\left\{  \phi,\psi\right\}
\rightarrow\left\{  \bar{\phi},\bar{\psi}\right\}  $ which diagonalizes the
kinetic metric depends on $\varepsilon^{-1}$, i.e. it is singular
transformation. Which means that the action (\ref{ac.08}) can not be written
in that of the Quintom model \cite{quin1,quin2}. Furthermore, the second fluid
$\psi$, is not a real fluid but it follows from the higher-order derivatives
of the field equations.

In particular, the reduction of the order of the field equations from a
fourth-order theory to a second-order, by using a Lagrange multiplier, it
means that in the same time the number of the dependent variables is
increased. An other analogue is the $f\left(  R\right)  $-gravity in the
metric formalism \cite{bb}. The last gravitational theory for nonlinear
$f\left(  R\right)  $ function is reduced with the use of the Lagrange
multiplier in a Brans-Dicke theory with vanished Brans-Dicke parameter
\cite{dilaton}, the so called O'Hanlon massive dilaton theory \cite{dil1}. For
other applications of Lagrange multipliers in different theories of gravity,
see \cite{lan1,lan2,lan3}, and for a discussion on the degrees of freedom in
alternative theories of gravity, see \cite{sott}.

In the following section we continue with the derivation of the field
equations in a spatially flat FLRW spacetime.

\subsection{Field equations in a FLRW background}

Let us consider a spatially flat FLRW spacetime with line element
\begin{equation}
ds^{2}=dt^{2}-a^{2}\left(  t\right)  \left(  dx^{2}+dy^{2}+dz^{2}\right)  ,
\label{frw.01}%
\end{equation}
and, with gravitational action (\ref{ac.08}), where the matter tensor $\bar
{T}_{\mu\nu}$ describes a dust fluid with energy density $\rho_{m}$. Moreover,
from the conservation law $\bar{T}_{~~~;\nu}^{\mu\nu}=0$ (Bianchi identity),
and, for the space (\ref{frw.01}), we have that $\rho_{m}=\rho_{m0}a^{-3}$,
where $\rho_{m0}$ is the present value of the energy density. From the
cosmological principle, we have that the fields $\left\{  \phi,\psi\right\}  $
inherit the symmetries of the metric (\ref{frw.01}), hence, $\phi=\phi\left(
t\right)  ,~\psi=\psi\left(  t\right)  $.

Therefore, for the space (\ref{frw.01}), the gravitation equations
(\ref{einstein}) follow from the following Lagrangian\footnote{In the
following, we consider $8\pi G=1.$}
\begin{equation}
L=3a\dot{a}^{2}-\frac{1}{2}a^{3}\dot{\phi}^{2}-2\varepsilon a^{3}\dot{\phi
}\dot{\psi}+a^{3}V\left(  \phi\right)  -\varepsilon a^{3}\psi^{2}+\rho_{m0}.
\label{ac.09}%
\end{equation}

Lagrangian (\ref{ac.09}) can be seen as the Lagrangian of a particle moving in
a Riemannian manifold with metric $\gamma_{ij},$ and effective potential
$V_{eff}=-\left(  a^{3}V\left(  \phi\right)  -\varepsilon a^{3}\psi^{2}%
+\rho_{m0}\right)  $, that is,
\begin{equation}
L\left(  y^{j},\dot{y}^{j}\right)  =\frac{1}{2}\gamma_{ij}\dot{y}^{i}\dot
{y}^{j}-V_{eff}\left(  y^{k}\right)  , \label{ac.10a}%
\end{equation}
where, $y^{i}=\left(  a,\phi,\psi\right)  $, and $\gamma_{ij}$ is given by the
following line element%
\begin{equation}
ds_{\gamma}^{2}=6ada^{2}-a^{3}d\phi^{2}-4\varepsilon a^{3}d\phi d\psi.
\label{ac.10b}%
\end{equation}

For the Euler-Lagrange equations of (\ref{ac.10a}) we have that these are
given by the following expression%
\begin{equation}
\gamma_{ik}\ddot{y}^{k}+\gamma_{ir}\Gamma_{jk}^{r}\dot{y}^{j}\dot{y}%
^{k}+\left(  V_{eff}\right)  _{,i}=0, \label{ac.10c}%
\end{equation}
where $\Gamma_{jr}^{k}$ are the Christoffel symbols of $\gamma_{ij}$, and the
nonzero Christoffel symbols are
\begin{equation}
\Gamma_{aa}^{a}=\frac{1}{2a}~,~\Gamma_{\phi\phi}^{a}=\frac{a}{4}%
~,~\Gamma_{\phi\psi}^{a}=\frac{\varepsilon}{2}a,
\end{equation}
and%
\begin{equation}
\Gamma_{a\phi}^{\phi}=\frac{3}{2a}~,~\Gamma_{a\psi}^{\psi}=\frac{3}{2a}.
\label{ac.10e}%
\end{equation}

Therefore from (\ref{ac.10c})-(\ref{ac.10e}), we have that the second
Friedmann's equation which is the Euler-Lagragne equation of (\ref{ac.09})
with respect to the scale factor, $a,$ is given as follows
\begin{equation}
\ddot{a}+\frac{1}{2a}\dot{a}^{2}+\frac{a}{4}\dot{\phi}^{2}+a\varepsilon
\dot{\phi}\dot{\psi}-\frac{1}{2}a\left(  V\left(  \phi\right)  -\varepsilon
\psi^{2}\right)  =0, \label{ac.11}%
\end{equation}
and the Klein-Gordon equations for the \textquotedblleft two
field\textquotedblright\ $\left\{  \phi,\psi\right\}  $, are
\begin{align}
\ddot{\phi}+\frac{3}{a}\dot{a}\dot{\phi}-\psi &  =0,\label{ac.12a}\\
\varepsilon\left(  \ddot{\psi}+\frac{3}{a}\dot{a}\dot{\psi}\right)  +\frac
{1}{2}\left(  \psi+V_{,\phi}\right)   &  =0, \label{ac.12b}%
\end{align}
which follow from the conservation law $T_{~~~;\nu}^{\mu\nu}\left(  \phi
,\psi\right)  =0.$ Furthermore, the first Friedmann's equation can be seen as
the Hamiltonian function which follows from Lagrangian (\ref{ac.09}), and it
is\footnote{Alternatively equation (\ref{ac.10}) can be derived from the
Euler-Lagrange equation with respect to a new variable, $N,$ which arises from
the lapsed time $dt=Nd\tau$, for instance see \cite{Ray}.},%
\begin{equation}
\rho_{m0}=3a\dot{a}^{2}-\frac{1}{2}a^{3}\dot{\phi}^{2}-2\varepsilon a^{3}%
\dot{\phi}\dot{\psi}-a^{3}V\left(  \phi\right)  +a^{3}\varepsilon\psi^{2}.
\label{ac.10}%
\end{equation}

Here, we would like to note that, the set of the field equations
(\ref{ac.10})-(\ref{ac.12b}) form a singular perturbation dynamical system.
Moreover, it is straightforward to prove that, equation (\ref{ac.12a}) is the
constrained equation $\psi=g^{\mu\nu}\phi_{;\mu\nu}$, and, if we replace
$\psi$ from (\ref{ac.12a}) in (\ref{ac.12b}), then we will have a fourth-order
equation, which is the modified Klein-Gordon equation for the field, $\phi$,
in GUP. Furthermore, in the limit in which $\varepsilon=0$, that is, the
deformation parameter $\beta$ is vanished and we have the classical
uncertainty principle, Lagrangian (\ref{ac.09}) is that of quintessence scalar
field with dust fluid, and in the same time equations (\ref{ac.11}%
)-(\ref{ac.12b}) reduced to the field equations for the last model.

An alternative way to write the field equations is,%
\begin{align}
&  3H^{2}=\rho_{GUP}+\rho_{m},\label{ac.14}\\
&  2\dot{H}+3H^{2}=-p_{GUP}, \label{ac.15}%
\end{align}
where $\rho_{GUP}=\rho_{\phi}+\rho_{C},~p_{GUP}=p_{\phi}+p_{C}$. Here,
$\rho_{\phi}$, $~p_{\phi}$ are the density and pressure of the usual scalar
field given by
\begin{equation}
\rho_{\phi}=\frac{1}{2}\dot{\phi}^{2}+V\left(  \phi\right)  ~,~p_{\phi}%
=\frac{1}{2}\dot{\phi}^{2}-V\left(  \phi\right)  , \label{ac.17}%
\end{equation}
and $\rho_{C}$, $p_{C}$ are the perturbation terms
\begin{equation}
\rho_{C}=\varepsilon\left(  2\dot{\phi}\dot{\psi}-\psi^{2}\right)
~,~p_{C}=\varepsilon\left(  2\dot{\phi}\dot{\psi}+\psi^{2}\right)  .
\label{ac.18}%
\end{equation}
However, the two fluids $\left\{  \rho_{\phi},\rho_{C}\right\}  $ are
interacting; furthermore, the field $\rho_{C}$,$~$i.e. equation (\ref{ac.18})
is not a real field, but it was introduced by the quantum corrections of GUP.

Finally, the effective EoS parameter for the scalar field which follows from
GUP is
\begin{equation}
w_{GUP}=\frac{\left(  \frac{1}{2}\dot{\phi}^{2}-V\left(  \phi\right)  \right)
+\varepsilon\left(  2\dot{\phi}\dot{\psi}+\psi^{2}\right)  }{\left(  \frac
{1}{2}\dot{\phi}^{2}+V\left(  \phi\right)  \right)  +\varepsilon\left(
2\dot{\phi}\dot{\psi}-\psi^{2}\right)  }. \label{ac.20}%
\end{equation}
Using the Taylor series expansion of equation (\ref{ac.20}) at $\varepsilon
\rightarrow0,$ we have $w_{GUP}=w_{\phi}+w^{\prime}$, where $w_{\phi}%
=\frac{p_{\phi}}{\rho_{\phi}},~$is the EoS parameter for the ``classical''
scalar field and
\begin{equation}
w^{\prime}=\frac{1}{\rho_{\phi}}\left(  p_{C}-w_{\phi}\rho_{C}\right)
~+O\left(  \varepsilon^{2}\right)  . \label{ac.20c}%
\end{equation}

In the limit~where $\dot{\phi}^{2}\ll V~$, which means that $w_{\phi}%
\simeq-{1}$, then~$w^{\prime}=\varepsilon\frac{4}{V}\dot{\phi}\dot{\psi}%
;~$therefore $w_{GUP}$ becomes
\begin{equation}
w_{GUP}\simeq-1+\frac{4\varepsilon}{V}\dot{\phi}\dot{\psi}\lesssim
-1,~\text{when}~~\dot{\phi}\dot{\psi}\leq0. \label{ac.22}%
\end{equation}
Hence, the EoS parameter $w_{GUP}$ can cross the phantom barrier provided when
$\dot{\phi}\dot{\psi}$ $<0,~$and, $\left\vert \dot{\psi}\right\vert >>1$.
Here, we would like to note that, since the field equations form a singular
perturbation system (slow-fast system), it is possible that $\left\vert
\dot{\psi}\right\vert >>1$, which means that either when the parameter
$\varepsilon$ is very small, \ the product $\left\vert \varepsilon\dot{\psi
}\right\vert \ $could be big enough in order the perturbation effects to be
measurable. In order to show this, in the following section we study the
existence of fixed points for the singular perturbation system in the slow
manifold. Finally, we perform numerical simulations for some well known scalar
field models, and, we show that, $w_{GUP}$ can cross the phantom divide line.

\section{Cosmological evolution}

\label{cosmoevol}

In this section, we consider the field equations (\ref{einstein}) where now
the energy momentum tensor $\bar{T}_{\mu\nu}$ is $\bar{T}_{\mu\nu}=\bar
{T}_{\mu\nu}^{\left(  m\right)  }+\bar{T}_{\mu\nu}^{\left(  r\right)  }%
~$where~$\bar{T}_{\mu\nu}^{\left(  m\right)  },$ $\bar{T}_{\mu\nu}^{\left(
r\right)  },$ are the energy-momentum tensors for a dust fluid~$\left(
w_{m}=0\right)  $ (dark matter), and a radiation fluid~$\left(  w_{r}=\frac
{1}{3}\right)  $ respectively.

The field equations (\ref{einstein}) for the spatially flat FLRW spacetime are%
\begin{align}
3H^{2}  &  =\rho_{\phi}+\rho_{m}+\rho_{r}+\varepsilon\left(  2\dot{\phi}%
\dot{\psi}-F\left(  \psi\right)  \right)  ,\label{dynamical1}\\
2\dot{H}+3H^{2}  &  =-p_{\phi}-p_{r}-\varepsilon\left(  2\dot{\phi}\dot{\psi
}+F\left(  \psi\right)  \right)  , \label{dynamical2}%
\end{align}
where $F\left(  \psi\right)  =\psi^{2}$.

We define the dimensionless variables \cite{copeland,fadragas}
\begin{align}
x_{1}  &  =\frac{\dot{\phi}}{\sqrt{6}H}~,~x_{2}=\frac{\sqrt{V\left(
\phi\right)  }}{\sqrt{3}H},~x_{3}=\frac{\sqrt{\rho_{r}}}{\sqrt{3}%
H},~\label{dyn.021}\\
y_{1}  &  =\frac{2\sqrt{6}}{3}\frac{\dot{\psi}}{H}~,~y_{2}=\frac{\sqrt{F}%
}{\sqrt{3}H}~~,~\Omega_{m}=\frac{\rho_{m}}{3H^{2}}. \label{dyn.022}%
\end{align}

In the new variables, the first Friedmann's equation (\ref{dynamical1}) can be
written as follows%
\begin{equation}
\Omega_{m}=1-\Omega_{GUP}-\Omega_{r}~, \label{dynamical1.1}%
\end{equation}
where now $\Omega_{r}=\left(  x_{3}\right)  ^{2},~$and $~\Omega_{GUP}=\left[
\left(  x_{1}\right)  ^{2}+\left(  x_{2}\right)  ^{2}\right]  +\varepsilon
\left[  x_{1}y_{1}-\left(  y_{2}\right)  ^{2}\right]  $.

Furthermore, the second Friedmann's equation (\ref{dynamical2}) takes the
form~$~\frac{\dot{H}}{H^{2}}=\frac{3}{2}\Xi\left(  \mathbf{x},\mathbf{y}%
,\varepsilon\right)  ,~$where%
\[
\Xi\left(  \mathbf{x},\mathbf{y},\varepsilon\right)  \equiv-1-\left[  \left(
x_{1}\right)  ^{2}-\left(  x_{2}\right)  ^{2}\right]  -\varepsilon\left[
x_{1}y_{1}+\left(  y_{2}\right)  ^{2}\right]  -\frac{1}{3}\left(
x_{3}\right)  ^{2},
\]
hence, the total EoS parameter is $w_{eff}=-1-\Xi\left(  \mathbf{x}%
,\mathbf{y},\varepsilon\right)  .$

The EoS parameter (\ref{ac.20}) for the scalar field in the new variables
becomes
\begin{equation}
w_{GUP}=\frac{\left[  \left(  x_{1}\right)  ^{2}-\left(  x_{2}\right)
^{2}\right]  +\varepsilon\left(  x_{1}y_{1}+\left(  y_{2}\right)  ^{2}\right)
}{\left[  \left(  x_{1}\right)  ^{2}+\left(  x_{2}\right)  ^{2}\right]
+\varepsilon\left(  x_{1}y_{1}-\left(  y_{2}\right)  ^{2}\right)  },
\label{eosd}%
\end{equation}
where in the limit $w_{\phi}\rightarrow-1,~$becomes
\begin{equation}
w_{GUP}=-1+2\varepsilon\frac{x_{1}y_{1}}{\left(  x_{2}\right)  ^{2}}+O\left(
\varepsilon^{2}\right)  . \label{eosd1}%
\end{equation}
\qquad\qquad\qquad

The derivation of the variables $\mathbf{x},\mathbf{y}$ with respect to the
lapse time $N=\ln a$ gives the following dynamical system%
\begin{equation}
\frac{d\mathbf{x}}{dN}=\mathbf{f}\left(  \mathbf{x,y},\lambda,\mu
,\varepsilon\right)  ~~,~\frac{d\lambda}{dN}=h_{1}\left(  \mathbf{x,y}%
,\lambda,\mu,\varepsilon\right)  , \label{dyn.01}%
\end{equation}%
\begin{equation}
\frac{dy_{2}}{dN}=g_{2}\left(  \mathbf{x,y},\lambda,\mu,\varepsilon\right)
~,~\frac{d\mu}{dN}=h_{2}\left(  \mathbf{x,y},\lambda,\mu,\varepsilon\right)  ,
\label{dyn.02}%
\end{equation}%
\begin{equation}
\varepsilon\frac{dy_{1}}{dN}=g_{1}\left(  \mathbf{x,y},\lambda,\mu
,\varepsilon\right)  , \label{dyn.03}%
\end{equation}
where the new variables $\lambda,\mu$ are $\lambda=\left(  -\frac{V_{,\phi}%
}{V}\right)  ~,~\mu=\left(  -\frac{1}{2}\frac{F_{,\psi}}{F}\right)  $, and,
the functions $\mathbf{f},\mathbf{g}~$and $\mathbf{h}$ are as follows:
\begin{align}
f_{1}  &  \equiv-3x_{1}-\frac{\sqrt{6}}{2}\mu\left(  y_{2}\right)  ^{2}%
-\frac{3}{2}x_{1}\Xi~.\label{dyn.05}\\
f_{2}  &  \equiv-\frac{\sqrt{6}}{2}\lambda x_{1}x_{2}-\frac{3}{2}x_{2}%
\Xi~.\label{dyn.06}\\
f_{3}  &  =-2x_{3}-\frac{3}{2}x_{3}\Xi~.\\
g_{1}  &  \equiv-3\varepsilon y_{1}+\sqrt{6}\lambda\left(  x_{2}\right)
^{2}+\sqrt{6}\mu\left(  y_{2}\right)  ^{2}-\frac{3}{2}\varepsilon y_{1}%
\Xi~.\label{dyn.07}\\
g_{2}  &  \equiv-\frac{\sqrt{6}}{4}\mu y_{1}y_{2}-\frac{3}{2}y_{2}%
\Xi~.\label{dyn.08}\\
h_{1}  &  \equiv-\sqrt{6}\lambda^{2}\left(  \Gamma_{1}-1\right)
x_{1}~;~\Gamma_{1}=\frac{V_{,\phi\phi}V}{V_{,\phi}^{2}}~.\label{dyn.09}\\
h_{2}  &  \equiv-\frac{\sqrt{6}}{2}\mu^{2}\left(  \Gamma_{2}-1\right)
y_{1}~;~\Gamma_{2}=\frac{F_{,\psi\psi}F}{F_{,\psi}^{2}}~. \label{dyn.10}%
\end{align}

\subsection{Slow-fast dynamical system}

\label{slowfast}

The dynamical system (\ref{dyn.01})-(\ref{dyn.03}) is a singular perturbation
system; specifically, it is a slow-fast system of first-order ordinary
differential equations, where $y_{1}$ is the fast variable
\cite{Tikhonov,Fen79,Fusco,Nils}. Under the transformation $\bar
{N}=\varepsilon N$, the dynamical system (\ref{dyn.01})-(\ref{dyn.03})
becomes
\begin{equation}
\frac{d\mathbf{x}}{d\bar{N}}=\varepsilon\mathbf{f}~,~\frac{d\lambda}{d\bar{N}%
}=\varepsilon h_{1}~,~\frac{dy_{1}}{d\bar{N}}=g_{1},~\frac{dy_{2}}{d\bar{N}%
}=\varepsilon g_{2}~,~\frac{d\mu}{d\bar{N}}=\varepsilon h_{2}~,
\label{dyn.0011}%
\end{equation}
which is a regular perturbation system \cite{Fen79}.\ The two dynamical
systems (\ref{dyn.01})-(\ref{dyn.03}) and (\ref{dyn.0011}) are equivalent but
in different \textquotedblleft time scales\textquotedblright.

In order to continue, we will use the Tikhonov's theorem \cite{Tikhonov,Nils},
and, we will study the fixed points of the dynamical system (\ref{dyn.01}%
)-(\ref{dyn.03}) in the slow manifold $g_{2}\left(  \mathbf{x,y},\lambda
,\mu,0\right)  =0$. This solution is called outer solution of the singular
perturbed system.

In the limit $\varepsilon=0$, the dynamical system is equivalent to the
algebraic dynamical system (\ref{dyn.01}), (\ref{dyn.02}) and $g_{2}\left(
\mathbf{x,y},\lambda,\mu,0\right)  =0;$ the last gives the slow manifold
$\lambda\left(  x_{2}\right)  ^{2}+\mu\left(  y_{2}\right)  ^{2}=0$.
Furthermore, when $\varepsilon=0$, from (\ref{dynamical1.1}), we have that
$\mathbf{xx}^{T}\leq1$; that is, for the variables $\mathbf{x~}$,
$\mathbf{x\in}\left[  -1, 1\right]  $ holds. Moreover, since $V\left(
\phi\right)  \geq0$, the parameter $x_{2}$ is constrained as follows:
$x_{2}\in\left[  0,1\right]  $.

We consider the exponential potential $V\left(  \phi\right)  =V_{0}%
e^{-\sigma\phi}$, where $\sigma$ is a constant different from zero, then
$\lambda=\sigma$; hence we have that $\Gamma_{1}=1,~\Gamma_{2}=\frac{1}{2}$.
The fixed points of the outer solution with the physical value of the problem
and the corresponding eigenvalues are given in table \ref{InnerSolution}.%

\begin{table}[tbp] \centering
\caption{Fixed points analysis of the evolution equations (\ref{dyn.01})-(\ref{dyn.03}) in the slow manifold for the exponential potential}%
\begin{tabular}
[c]{ccccccccccc}\hline\hline
\textbf{Point} & $\left(  \mathbf{x}_{1},\mathbf{x}_{2},\mathbf{x}_{3}\right)
$ & $\left(  \mathbf{y}_{1},\mathbf{y}_{2}\right)  $ & $\mathbf{\mu}$ &
\textbf{Eigenvalues} & \textbf{Stability} & $\mathbf{\Omega}_{m}$ &
$\mathbf{\Omega}_{r}$ & $\mathbf{\Omega}_{GUP}$ & $\mathbf{w}_{eff}$ &
$\mathbf{w}_{GUP}$\\\hline
$O$ & $\left(  0,0,0\right)  $ & $\left(  0,0\right)  $ & $\mu$ & $\frac{3}%
{2}\left(  -1,1,1,0\right)  $ & Unstable & $1$ & $0$ & $0$ & $0$ & $\nexists
$\\
$O^{\ast}$ & $\left(  0,0,0\right)  $ & $\left(  y_{1},0\right)  $ & $0$ &
$\frac{3}{2}\left(  -1,1,1,0\right)  $ & Unstable & $1$ & $0$ & $0$ & $0$ &
$\nexists$\\
$A_{\pm}$ & $\left(  \pm1,0,0\right)  $ & $\left(  0,0\right)  $ & $\mu$ &
$\left(  3,3,3\mp\frac{3\lambda}{\sqrt{6}},0\right)  $ & Unstable & $0$ & $0$
& $1$ & $1$ & $1$\\
$A_{\pm}^{\ast}$ & $\left(  \pm1,0,0\right)  $ & $\left(  y_{1},0\right)  $ &
$0$ & $\left(  3,3,3\mp\frac{3\lambda}{\sqrt{6}},0\right)  $ & Unstable & $0$
& $0$ & $1$ & $1$ & $1$\\
$R_{\pm}$ & $\left(  0,0,\pm1\right)  $ & $\left(  0,0\right)  $ & $\mu$ &
$\left(  -1,2,2,0\right)  $ & Unstable & $0$ & $1$ & $0$ & $\frac{1}{3}$ &
$\nexists$\\
$R_{\pm}^{\ast}$ & $\left(  0,0,\pm1\right)  $ & $\left(  y_{1},0\right)  $ &
$0$ & $\left(  -1,2,2,0\right)  $ & Unstable & $0$ & $1$ & $0$ & $\frac{1}{3}$
& $\nexists$\\\hline\hline
\end{tabular}
\label{InnerSolution}%
\end{table}%

The fixed points$~$of the table \ref{InnerSolution} exist for all values of
the parameter $\lambda.$ Points $O,O^{\ast}$ correspond to the matter epoch,
where the total EoS parameter is $w_{tot}=0$. At the points $A_{\pm},~A_{\pm
}^{\ast}$, the universe is dominated by the kinetic energy of the scalar field
$\left(  \Omega_{\phi}=1,~V\left(  \phi\right)  =0\right)  ,$ which means that
the scalar field acts as a stiff fluid, i.e. $\rho_{\phi}=\rho_{\phi0}a^{-6}%
~$, this solution corresponds to a non-accelerated universe. The fixed points
$R_{\pm},R_{\pm}^{\ast}$ correspond to the radiation epoch, where $\Omega
_{r}=1$ and $w_{tot}=\frac{1}{3}$. Finally, the linearized system
(\ref{dyn.01}), (\ref{dyn.02}) at the fixed points admits positive
eigenvalues; that is, all the fixed points are unstable.

However, in contrary to the \textquotedblleft classical\textquotedblright%
\ minimal coupled scalar field cosmology, there are not the scalar field
dominated and the tracker solutions \cite{Ame10}. We would like to note that,
this analysis holds for very small values of the parameter $\Delta N=N-N_{0}$;
where $N_{0}$ is the value which we consider as a starting point.

When $\varepsilon=0,$ the regular perturbation system (\ref{dyn.0011}) gives
that the parameters $\mathbf{x},\lambda,y_{2},\mu$ are constants (that holds
for any potential), and, the only non-constant equation is
\begin{equation}
\frac{dy_{1}}{d\bar{N}}=\sqrt{6}\lambda\left(  x_{2}\right)  ^{2}+\sqrt{6}%
\mu\left(  y_{2}\right)  ^{2}=\alpha=\mbox{const}, \label{dyn.12}%
\end{equation}
that is, $y_{1}\left(  \bar{N}\right)  =$ $\alpha\bar{N}+\beta,$ where
$\alpha,\beta\in%
\mathbb{R}
$. However, since $\varepsilon=0$, the cosmological parameters are independent
on $y_{1}$; that is, for very large values of the parameter $\left(
N-N_{0}\right)  =\ln a$, the cosmological parameters $\Omega_{m},~\Omega_{r},$
etc. will be constants.

In order to understand better the effects of the GUP in the dark energy
models, we perform numerical simulations for the dynamical system
(\ref{dyn.01})-(\ref{dyn.03}) for the power law model $V\left(  \phi\right)
=V_{0}\phi^{-1}$ \cite{Ame10}, and, the hyperbolic model $V\left(
\phi\right)  =V_{0}\left(  \cosh\left(  \sigma\phi\right)  -1\right)  ^{p}$
\cite{Sahni}.

\begin{figure}[ptb]
\includegraphics[height=8cm]{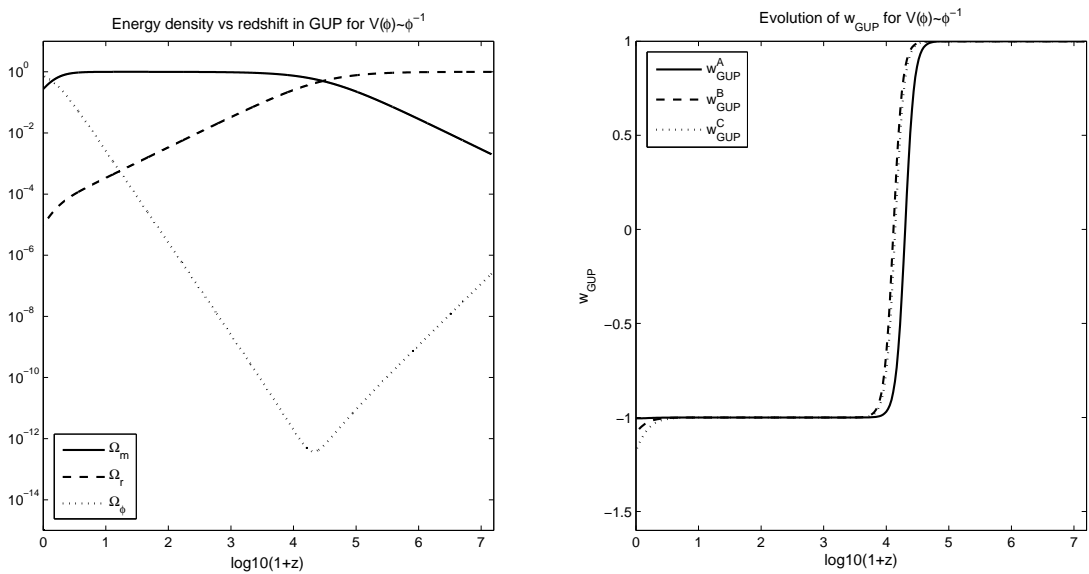}
\caption{Numerical solution of the field equation in the dimensionless
variables for minimally coupled scalar field cosmology modified by the GUP for
the power law potential $V\left(  \phi\right)  =V_{0}\phi^{-1}$. The left fig.
is the evolution of the energy density of the $\Omega_{m}~$(solid
line),~$\Omega_{r}$~(dashed line) and $\Omega_{\phi}~$(dot line). For the
numerical simulation we start from points in the radiation epoch,
$\log10\left(  1+z\right)  =7.21$. The right fig. is the evolution of the
scalar field equation of state parameter (EoS) for the same model. For the
numerical integration, we have considered $\beta_{0}=1$.}%
\label{figgupamen1}%
\end{figure}

In fig. \ref{figgupamen1}, we give the numerical solutions of the energy
density evolution, and of the EoS parameter $w_{GUP}$ for the the potential
$V\left(  \phi\right)  =V_{0}\phi^{-1}$. For the non-modified scalar field,
the same plots can be found in \cite{Ame10}. We observe that in contrary with
the \textquotedblleft classical\textquotedblright\ quintessence scalar field,
where in the tracking solution, the EoS parameter has the value $w_{\phi
}=-\frac{2}{3},$ whereas the $w_{GUP}$ can cross the phantom divide line;
however, the energy density evolution of the two models has similar behavior.

\begin{figure}[ptb]
\includegraphics[height=8cm]{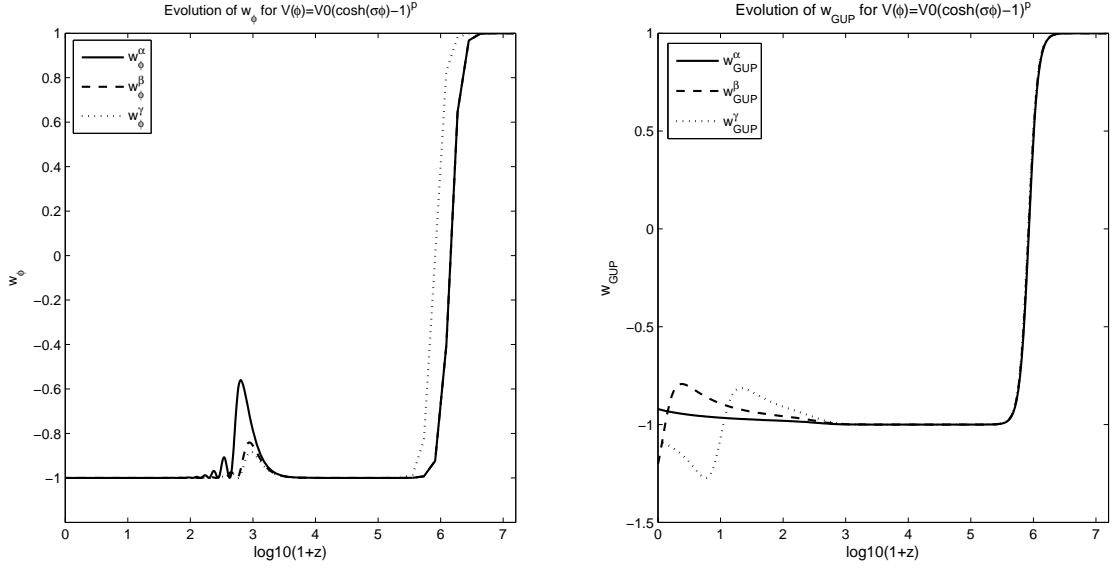}
\caption{Numerical solution of the EoS parameter $w_{\phi}$ and $w_{GUP}$ for
minimally coupled scalar field cosmology for the hyperbolic potential
$V\left(  \phi\right)  =V_{0}\left(  \cosh\left(  \sigma\phi\right)
-1\right)  ^{p}$ with $p=0.2$, and, $\sigma=20$ \cite{Sahni}. The left fig.
describes the evolution of the EoS parameter $w_{\phi}~$for the ``classical''
scalar field, whereas the right fig. is for the modified EoS parameter
$w_{GUP}$. \ For the numerical simulation, we start from points in the
radiation epoch, $\log10\left(  1+z\right)  =7.21$, and, we consider
$\beta_{0}=1.$}%
\label{sahnieos}%
\end{figure}

Furthermore, in fig. \ref{sahnieos}, we compare the evolution of the EoS
parameters $w_{\phi}$ and for the modified EoS parameter $w_{GUP}$ for the
hyperbolic model $V\left(  \phi\right)  =V_{0}\left(  \cosh\left(  \sigma
\phi\right)  -1\right)  ^{p}$. We observe that the scalar field modified by
GUP $\ $can cross the phantom divide line, i.e. $w_{GUP}<-1$.

\begin{figure}[ptb]
\includegraphics[height=8cm]{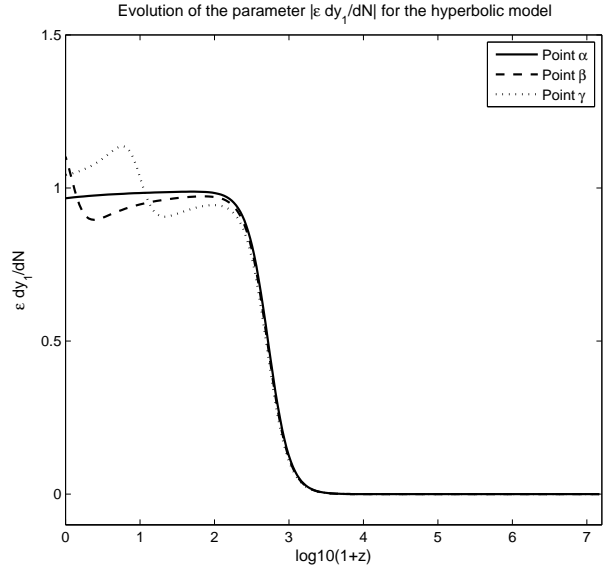}
\caption{Numerical solution of the parameter $\varepsilon dy_{1}/dN$ (equation
(\ref{dyn.03}) ) for the hyperbolic model of fig.\ref{sahnieos}. We observe
that, for large values of $\Delta N$, the fast variable $y_{1}$ take values
$y_{1}\simeq\varepsilon^{-1}$, and, when $\varepsilon dy_{1}/dN>1$, the
$w_{GUP}$ crosses the phantom divide line, which means that the term
$\varepsilon y_{1}$ of equation (\ref{eosd1}) is measurable. For the numerical
integration, we have considered $\beta_{0}=1$. However, a similar behavior
holds, and for different values of $\beta_{0}$.}%
\label{fastvar}%
\end{figure}

Finally, in order to understand the behavior of the slow-fast system
(\ref{dyn.01})-(\ref{dyn.03}), in fig. \ref{fastvar}, we give the evolution of
the parameter $\left\vert \varepsilon\frac{dy_{1}}{dN}\right\vert $ for the
hyperbolic potential. From the fig. \ref{fastvar}, we observe that for large
$\Delta N$ parameter, $y_{1}$ is fast, i.e. $y_{1}\simeq\varepsilon^{-1}$,
which means that the term $\varepsilon y_{1}~$in equation (\ref{eosd1}) can be
measurable. \ For the numerical solutions, in figs. \ref{figgupamen1},
\ref{sahnieos} and \ref{fastvar}, we selected $\varepsilon=10^{-70}$, which
corresponds to the natural value of the parameter $\beta_{0}\,,$ i.e.
$\beta_{0}\simeq1$, However, we would like to remark that the behavior of the
solutions is similar and for smaller values of the parameter $\varepsilon$,
i.e. $\beta_{0}$; since when the parameter $\varepsilon$ decrease, the fast
variable $y_{1}$ becomes \textquotedblleft faster\textquotedblright.

\section{Conclusions}

\label{Discu}

In this work, we have considered a modified quintessence scalar field
Lagrangian which follows from the GUP. The modified Klein-Gordon equation for
the scalar field admits a fourth-order perturbation term which follows from
the quantum corrections. Furthermore, we have considered the gravitational
action, and, with the use of Lagrange multiplier, we rewrote the gravitational
action as that of a two scalar fields model with a mixed kinetic term; we
remark that, the second field was introduced by the quantum corrections of GUP
and it is not a real field. The proposed model, i.e. the action integral
(\ref{ac.08}), it can be compared with that of a quintom model in which two
fields take place. However, in our consideration there exist only one field
which satisfies a fourth-order Klein-Gordon equation and in general it is
different from the quintom models, moreover the coordinate transformation
which diagonalize the kinetic term, which is defined by the field $\phi$, and
the Lagrange multiplier $\psi$, is a singular transformation; which means that
the inverse transformation can not be defined.

Consider now the action integral of a quintom model with mixed kinetic term,
which has been introduced in \cite{Saridakis1},%
\begin{equation}
S_{mix}=\int d^{4}x\left[  R-\frac{\phi_{0}}{2}g^{\mu\nu}\phi_{;\mu}\phi
_{;\nu}+\frac{\psi_{0}}{2}g^{\mu\nu}\psi_{;\mu}\psi_{;\nu}-\frac{\alpha}%
{2}g^{\mu\nu}\phi_{;\mu}\psi_{;\nu}+V_{1}\left(  \phi\right)  +V_{2}\left(
\psi\right)  \right]  \label{sf.01}%
\end{equation}
where $\phi_{0},\psi_{0,}\alpha$ are constants\footnote{In \cite{Saridakis1},
$\phi_{0}=\psi_{0}=1$}, then in comparison with the action integral
(\ref{ac.08}), it can be seen that, $\psi_{0}=0$, $V_{2}\left(  \psi\right)  $
is fixed, which follows from the Lagrange multiplier, and the constant,
$\alpha$, is related with the deformation parameter of the GUP, i.e.
$\alpha\simeq\beta$. Hence a quintom scenario with mixed terms in which the
parameter, $\alpha$, is small, can arise form the GUP.

In the case of a spatially flat FLRW spacetime, we derived the field equations
and we wrote the modified cosmological parameters. We show that the EoS
parameter $w_{GUP}$ for the scalar field has not a minimum boundary; that is,
it is possible to cross the phantom divide line, such as, $w_{GUP}<-1$.
Moreover, in order to study the cosmological evolution of the model, we wrote
the field equations in the dimensionless variables, and, in the lapse time
$N=\ln a$. Since the dynamical system is a singular perturbation system we
studied the existence of fixed points in the slow manifold. We have shown
that, for an exponential potential, the outer solution of the field equations
admits only unstable fixed points, which correspond to the matter dominated
era, the radiation epoch, and, to the solution where the universe is dominated
by the kinetic part of the scalar field. For \textquotedblleft
large\textquotedblright\ values of the lapse time, we demonstrate that the
cosmological parameters are constants.

Finally, we have performed numerical simulations of the field equations for
two well known potentials $V\left(  \phi\right)  ;$ in both cases, the EoS
parameter $w_{GUP}$ can take values lower than minus one. That result is
important, since we can have a phantom behavior which follows for the quantum
corrections of a quintessence scalar field model.

\begin{acknowledgments}
The authors would like to thank the anonymous referees for various comments
and suggestions which improved the quality of the present work. AP
acknowledges financial support of INFN. S. Pan acknowledges CSIR, Govt. of
India for research grants through SRF scheme (File No: 09/096(0749)/2012-EMR-I).
\end{acknowledgments}

\end{document}